\documentclass[pre,twocolumn,aps,eqsecnum]{revtex4}
\usepackage{epsfig}

\begin{document}

\title{Statistical Thermodynamics of Strain Hardening in Polycrystalline Solids}

\author{J.S. Langer}
\affiliation{Department of Physics, University of California, Santa Barbara, CA  93106-9530}

\date{\today}

\begin{abstract}
This paper starts with a systematic rederivation of the statistical thermodynamic equations of motion for dislocation-mediated plasticity proposed in 2010 by Langer, Bouchbinder and Lookman. It then uses that theory to explain the anomalous rate-hardening behavior reported in 1988 by Follansbee and Kocks, and to explore the relation between hardening rate and grain size reported in 1995 by Meyers et al.  A central theme is the need for physics-based, nonequilibrium analyses in developing predictive theories of the strength of polycrystalline materials.  

\end{abstract}

\maketitle

\section{Introduction}
\label{Intro}

In an earlier publication \cite{LBL-10} (LBL), Bouchbinder, Lookman, and I proposed a statistical-thermodynamic framework for studying dislocation-mediated plasticity in polycrystalline solids.  Our purpose was to replace, or to find a firmer basis for, the largely phenomenological equations of motion that have been used for many decades in attempts to describe the dynamic behavior of these materials. We succeeded in computing the measured flow stress for Cu over fifteen decades of strain rate, and for temperatures between room temperature and about one third of the melting temperature, using just a few plausible, physics-based parameters.  We also computed stress-strain -- i.e. strain-hardening -- curves as functions of strain rate and temperature.  The success of these efforts has given some credibility to our first-principles reformulation of dislocation theory.  To test its limits of validity, we need next to look at specific phenomena and, more generally, to make closer contact with existing knowledge in this field. 

The LBL analysis is based on fundamental principles of statistical physics that, as pointed out in the introductory sections of that paper, are almost completely ignored in conventional dislocation theories. It seems obvious, however, that complex behaviors involving chaotic motions of very large numbers of entangled dislocations can be described only in statistical terms.  The equations of motion for these systems must necessarily be consistent with the laws of thermodynamics, especially the requirements of energy conservation and non-decreasing entropy.  Thus, these equations  must describe the flow of energy and entropy through driven, nonequilibrium systems.  In LBL, they are expressed in terms of two dynamical variables: the density of dislocations $\rho$, and an effective temperature $\chi$ that describes the system's state of configurational disorder.  These two variables determine responses to external driving forces, and carry the memory of earlier deformations.  

At the next level of specificity, the LBL analysis assumes that plastic deformation in polycrystalline solids is caused by the motions of line defects, i.e. dislocations, and that these motions are determined primarily by the rates at which dislocation lines become unpinned from each other under the influence of elastic stresses and thermal fluctuations.  LBL argued that pinning times are generally very much longer than the times taken by dislocations to move from one pinning site to another, and therefore neglected viscous forces affecting the motions of unpinned dislocations. Moreover, LBL paid no attention to details such as the distinction between ``edge'' and ``screw'' dislocations, or ``cross-slip,'' or ``stacking faults,'' or the like. Instead, it was assumed that such details would become relevant if and when it was necessary to make first-principles estimates of quantities, such as rate factors, that appear in the equations of motion for $\rho$ and $\chi$.  The latter details do not appear in this paper either.  However, ``grain boundaries''  appear in important ways. 

In what follows, I use the LBL statistical-thermodynamic framework to look at two long-standing puzzles in solid mechanics, neither of which has proved amenable to the conventional phenomenological analyses.  First, I discuss the abrupt upturn in stress that has been observed in Cu for small values of the strain but at large values of the strain rate.\cite{FOLLANSBEE-KOCKS-88} Second, I discuss what I believe is a related observation that strain hardening curves for Cu sharpen abruptly  when the grain size becomes sufficiently small.\cite{MEYERSetal-95}  Understanding these phenomena illustrates how specific mechanisms can be incorporated into the LBL theory where necessary.  It also illustrates how certain aspects of the thermodynamic theory -- not fully developed in LBL -- become relevant to practical applicatons.  Therefore, in what follows, I start by reviewing the thermodynamic basis of the LBL analysis.  

\section{Thermodynamic Equations of Motion}
\label{thermo}

This thermodynamic analysis has been discussed in detail in earlier publications devoted primarily to theories of plasticity in amorphous materials. For example, see \cite{FL-11,BL-I-II-III-09}. For both the amorphous and polycrystalline cases, the analysis starts by dividing the system into  configurational and kinetic-vibrational subsystems.  The configurational degrees of freedom describe the relatively slow,i.e. infrequent, atomic rearrangements that are associated with irreversible plastic deformation; the kinetic-vibrational degrees of freedom describe the fast thermal and vibrational motions of the atoms.

For the polycrystalline case, it is useful to think of a slab of material lying in the plane of an applied shear stress. The dislocations oriented perpendicular to this plane are driven by the stress to move through a ``forest'' of dislocations lying primarily in the plane, thus producing shear flow.  Let the macroscopic area of this slab be $A$, and let its thickness be a characteristic dislocation length, say, $L$.  

The total internal energy of this system is
\begin{equation}
U_{total} = U_C(S_C, \rho) + U_R(S_R).
\end{equation}
Here, $U_C(S_C, \rho)$ is the configurational energy of a polycrystalline material containing dislocations, grain boundaries, and other structural irregularities.  $\rho$ is the areal density of dislocations or, alternatively, the total length of dislocation lines per unit volume. $S_C(U_C,\rho)$ is the entropy of the configurational subsystem computed by counting the number of configurations at fixed values of $U_C$ and $\rho$. $U_R(S_R)$ is the kinetic-vibrational energy of this system, whose entropy is $S_R$.  The kinetic-vibrational subsystem serves as a thermal reservoir.  Its temperature, $k_B T = \theta = \partial U_R/\partial S_R$, is assumed to be fixed.   The effective temperature of the configurational subsystem, 
\begin{equation}
\chi = \left({\partial U_C\over\partial S_C}\right)_{\rho},
\end{equation}
plays a central role in this analysis.  

Assume that we can write 
\begin{equation}
\label{UC}
U_C(S_C, \rho) = U_0(\rho) + U_1(S_1),
\end{equation}
and, correspondingly, 
\begin{equation}
\label{SC}
S_C(U_C,\rho)= S_0(\rho) + S_1(U_1),
\end{equation}
where, $U_1$ and $S_1$ are, respectively, the energy and the entropy of all the configurational degrees of freedom other than those associated with the dislocations.  

$U_0(\rho)$ is the dislocation energy which, for present purposes, I write simply in the form 
\begin{equation}
\label{Urho}
U_0(\rho)= A\,\rho\,e_D;~~~~e_D = L\,\gamma_D,
\end{equation}
where $e_D$ is the energy per dislocation and $\gamma_D$ is the dislocation energy per unit length.  An implicit assumption here is that dislocations of opposite signs are present in equal numbers, thus minimizing the elastic energy.  I also assume that the residual elastic energy, of the order of $\rho\,\ln(\rho)$, is included approximately in $e_D$, but is too slowly varying a function of $\rho$ to be needed for present purposes.  It may eventually be needed for studying spatially varying dislocation patterns.  Finally, in Eq.(\ref{SC}), the entropy of the dislocations, $S_0(\rho) $, can be written approximately in the form
\begin{equation}
\label{Srho}
S_0(\rho) \approx -\,A\,\rho\,\ln(a^2\,\rho)+ A\,\rho ~~~{\rm for}~~ a^2\,\rho \ll 1,
\end{equation}
where $a$ is a length of the order of the atomic spacing.

The usual thermodynamic analysis for this system goes as follows.  The first law is:
\begin{eqnarray}
\label{firstlaw}
\nonumber
\dot U_{total}&=& V\,\sigma\,\dot\epsilon^{pl} = \dot U_C + \dot U_R\cr\\ &=& \chi\,\dot S_C + \left({\partial U_C\over \partial \rho}\right)_{S_C} \dot \rho + \theta\,\dot S_R,
\end{eqnarray}
where $V = LA$ is the volume, $\sigma$ is the shear stress, and $\dot\epsilon^{pl}$ is the plastic shear rate. (Variations of the reversible elastic energy cancel out of this equation. For example, see \cite{BL-I-II-III-09}.) Use Eq.(\ref{firstlaw}) to evaluate $\dot S_C$, and write the second law in the form
\begin{equation}
\label{dotS}
\dot S_C + \dot S_R = {1\over \chi}\,{\cal W} +\left(1 - {\theta\over \chi}\right)\dot S_R \ge 0,
\end{equation}
where
\begin{equation}
\label{W} 
{\cal W} = V\,\sigma\,\dot\epsilon^{pl} - \left({\partial U_C\over \partial \rho}\right)_{S_C}\dot \rho 
\end{equation}
is the difference between the power delivered to the system and the rate at which energy is stored in the form of dislocations.

Equation (\ref{dotS}) is the sum of independent inequalities that, according to an argument originally due to Coleman and Noll, must be satified separately.  Non-negativity of the term proportional to $\dot S_R$ implies that the heat flux $Q$, defined here to be positive when heat is flowing (as expected) from the configurational subsystem into the thermal reservoir, is
\begin{equation}
\label{Q}
Q = \theta \dot S_R = {\cal K}\,(\chi - \theta),
\end{equation}
where ${\cal K}$ is a non-negative thermal transport coefficient.  

For present purposes, assume that the mechanical power, $V\,\sigma\,\dot \epsilon^{pl}$, is always positive.  Therefore, the remaining inequality is
\begin{equation}
\label{F'neg}
\left({\partial U_C\over \partial \rho}\right)_{S_C}\,\dot \rho \le 0.
\end{equation}
Use Eqs. (\ref{UC}) and (\ref{SC}) to write $U_C = U_0 + U_1(S_C - S_0)$, so that 
\begin{eqnarray}
\nonumber
\left({\partial U_C\over \partial \rho}\right)_{S_C} &=& {\partial U_0\over \partial \rho} - \chi\,{\partial S_0\over \partial \rho} = {\partial F\over \partial \rho};\cr \\ F(\rho) &\equiv& U_0(\rho) - \chi\,S_0(\rho).
\end{eqnarray}
Equation (\ref{F'neg}) is satisfied by writing an equation of motion for $\rho$ in the form
\begin{equation}
\label{dotrho}
\dot \rho = -{\cal M}\,{\partial F\over \partial \rho},
\end{equation}
where ${\cal M}$ is a non-negative rate factor.  Next note that Eqs.(\ref{Urho}) and (\ref{Srho}) imply that $\partial F/\partial \rho = 0$ when $\rho = \rho_{ss}(\chi)$, where 
\begin{equation}
\rho_{ss}(\chi)= {1\over a^2}\,e^{-e_D/\chi} .
\end{equation}
It is simplest to rewrite Eq.(\ref{dotrho}) in the linearized form
\begin{equation}
\label{dotrho2}
\dot \rho = \tilde {\cal M}\,\left[1- {\rho \over \rho_{ss}(\chi)}\right].
\end{equation} 
The factor $\tilde {\cal M}$ must be proportional to the power per unit volume $\sigma\,\dot \epsilon^{pl}$, which is the only scalar rate in the problem.  It has the dimensions of energy per unit volume per unit time. The left-hand side of Eq.(\ref{dotrho2}) has the dimensions of length (of dislocations) per unit volume per unit time.  Thus, writing this equation in the form
\begin{equation}
\label{dotrho3}
\dot \rho = \kappa_{\rho}\,{\sigma\,\dot \epsilon^{pl}\over \gamma_D}\,\left[1- {\rho \over \rho_{ss}(\chi)}\right]
\end{equation}
is dimensionally correct and identifies the dimensionless factor $\kappa_{\rho}$ as the fraction of the input power that is converted into dislocations.  The second term on the right-hand side of Eq.(\ref{dotrho3}) can then be interpreted as the rate at which dislocations are annihilated as required by the second law.  

Having derived an equation of motion for $\rho$, return now to Eq.(\ref{firstlaw}) and rewrite this first-law equation in a form suitable for deriving an equation of motion for $\chi$:
\begin{equation}
\label{firstlaw2}
 \chi\,\dot S_C = V\,\sigma\,\dot \epsilon^{pl} - \left({\partial U_C\over \partial \rho}\right)_{S_C}\dot \rho -Q.
\end{equation}
Use the decompositions in Eqs.(\ref{UC}) and (\ref{SC}) to write the left-hand side as
\begin{equation}
\chi\,\dot S_C = \chi\, {\partial S_1\over \partial \chi}\, \dot \chi + \chi {\partial S_0\over \partial \rho}\,\dot \rho \equiv V\,c_{e\!f\!f}\,\dot \chi + \chi {\partial S_0\over \partial \rho}\,\dot \rho,
\end{equation}
which defines the effective specific heat $c_{e\!f\!f}$.  Next, make a similar expansion of the right-hand side of Eq.(\ref{firstlaw2}), and note that the term proportional to $\partial S_0/\partial \rho$ cancels out, leaving
\begin{equation}
\label{firstlaw3}
V\,c_{e\!f\!f}\,\dot \chi = V\,\sigma\,\dot \epsilon^{pl}  - {\partial U_0\over \partial \rho}\,\dot \rho -Q.
\end{equation}
In most STZ papers, my colleagues and I have neglected the analog of the term proportional to $\dot \rho$ on the right-hand side of Eq.(\ref{firstlaw3}), because the energy content of STZ's is negligible in comparison to that of all the other configurational degrees of freedom.  Here, however, it appears that a non-negligible fraction of the configurational energy may be stored in the dislocations.  

As argued, for example, in \cite{JSL-MANNING-TEFF-07}, the steady-state value of the effective temperature, say $\chi_{ss}$, must be a function of only the strain rate $\dot\epsilon$.  (In steady state, the total strain rate $\dot\epsilon$ is the same as the plastic strain rate $\dot\epsilon^{pl}$.) In other words, the steady state of disorder can depend only on the rate at which the system is being ``stirred'' by shearing.  Moreover, we argued that $\chi_{ss}$ must go to a non-negative constant, say $\chi_0$, in the limit of vanishing $\dot\epsilon$.  If $\dot\epsilon$ is slower than internal relaxation rates, then the steady state of disorder is determined only by the number of atomic rearrangements that are driven by the external forces, and not by the rate at which they occur.  In LBL, we pointed out that the rapid rise in the stress seen at strain rates comparable to atomic vibration frequencies can be interpreted as a rapid rise in $\chi$ when the system is being driven too fast for it to relax between rearrangement events.  In the situations to be considered here, however, the strain rates are not so large, and therefore I consider only cases where $\chi_{ss} = \chi_0$.

To use this observation in Eq.(\ref{firstlaw3}), note that $\chi$ is comparable to the mesoscopically large energy $e_D$, so that $\chi \gg \theta$ and $Q \approx {\cal K}\,\chi$ in Eq.(\ref{Q}). The requirement that $\chi_{ss} = \chi_0$ tells us that ${\cal K} = V\,\sigma\,\dot \epsilon^{pl}/\chi_0$, so that, with Eq.(\ref{Urho}), the equation of motion for $\chi$ becomes
\begin{equation}
\label{firstlaw4}
c_{e\!f\!f}\,\dot \chi = \sigma\,\dot \epsilon^{pl}\,\left[1 - {\chi\over \chi_0}\right]  - \gamma_D\,\dot \rho .
\end{equation}

In summary, the equations of motion for $\rho$ and $\chi$, Eqs.(\ref{dotrho3}) and (\ref{firstlaw4}) respectively, have been derived here using only basic principles of statistical thermodynamics and dimensional arguments. Equation (\ref{dotrho3}) describes the flow of energy through the system of dislocations, as constrained by the second law of thermodynamics.  Equation (\ref{firstlaw4}) describes the flow of entropy; it is a restatement of the first law.  

\section{Dynamics}
\label{Dynamics}

To complete this theory, we need relationships between the plastic strain rate $\dot\epsilon^{pl}$, the total strain rate $\dot\epsilon$, and the shear stress $\sigma$.  Start with the assumption that the elastic and plastic strain rates are simply additive quantities, so that 
\begin{equation}
\label{dotsigma}
\dot\sigma = \mu\,(\dot\epsilon - \dot\epsilon^{pl}),
\end{equation}
where $\mu$ is the shear modulus. The expression for the plastic strain rate is based on the Orowan relation:
\begin{equation}
\label{Orowan}
\dot\epsilon^{pl} = \rho\,b\,v,
\end{equation}
where $b$ is the magnitude of the Burgers vector and $v$ is the average speed at which dislocations move across the system.  That is, 
\begin{equation} 
v = {\ell\over \tau_P(\sigma)},
\end{equation}
where $\ell = 1/\sqrt{\rho}$ is the average spacing between dislocations, and $1/\tau_P(\sigma)$ is the depinning rate. As in LBL, assume that depinning is a thermally activated process with a stress-dependent barrier of the form
\begin{equation}
\label{UP}
U_P(\sigma) = k_B T_P\,e^{-\sigma/\sigma_T}.
\end{equation}
Here, the the height of the unstressed barrier is defined to be $k_B T_P$, and the characteristic depinning stress $\sigma_T$ is the Taylor stress,
\begin{equation}
\sigma_T = \mu_T\,b\,\sqrt{\rho},
\end{equation}
where $\mu_T$ is a reduced shear modulus of the order of $\mu/30$.  There is nothing sacrosanct about the exponential function in Eq.(\ref{UP}). A linear approximation would be satisfactory, because the ratio $\sigma/\sigma_T$ turns out to vary by not much more than a factor of two in the experiments to be discussed here. 

With these ingredients, the depinning rate is
\begin{equation}
  {1\over \tau_P(\sigma)} = {1\over \tau_0}\,f_P(\sigma),
\end{equation}
where 
\begin{equation}
f_P(\sigma)=\exp\,\left(-\,{T_P\over T}\,e^{-\sigma/\sigma_T}\right),
\end{equation}
and $\tau_0$ is an atomic time scale, set equal to $10^{-12}$ seconds throughout this analysis. It is convenient to use $\tau_0$ to define a dimensionless plastic strain rate, which, according to Eq.(\ref{Orowan}), is
\begin{equation}
\label{q-sigma}
q(\sigma,\tilde\rho) \equiv \tau_0\,\dot\epsilon^{pl} = \sqrt{\tilde\rho}\,\bigl[f_P(\sigma)-f_P(-\sigma)\bigr].
\end{equation}
where $\tilde\rho \equiv b^2\rho$. Antisymmetry on the right-hand side of this equation is formally required; but the second term in the square brackets is negligible for positive stresses in or above the strain-hardening regime.  Dropping that term, and setting $\sigma_T = \mu_T\sqrt{\tilde\rho}$, we can solve Eq.(\ref{q-sigma}) explicitly for the stress as a function of the strain-rate:
\begin{equation}
\label{nudef}
{\sigma\over \sigma_T} = \ln\,\left({T_P\over T}\right) - \ln\,\left[\ln\,\left({\sqrt{\tilde\rho}\over q}\right)\right]\equiv \nu(T,\tilde\rho,q).
\end{equation}

It is useful to non-dimensionalize these equations, and to replace the time by the strain $\epsilon$ as the independent variable, assuming a constant total strain rate $\dot\epsilon$.  Let $a = b$ in the normalization of $\rho$ given in Eq.(\ref{Srho}) and in the definition of $\tilde\rho$ in Eq.(\ref{q-sigma}).  (As shown in LBL, this is equivalent to making a small change in the time scale $\tau_0$.)   Define $\tilde\chi \equiv \chi/e_D$.  Also, let $\tau_0\,\dot\epsilon \equiv q_0$.  Then, Eqs.(\ref{dotsigma}), (\ref{dotrho3}), and (\ref{firstlaw4}) become, respectively,
\begin{equation}
\label{dotsigma2}
{d\sigma\over d \epsilon} = \mu\,\left[1 - {q(\sigma,\tilde\rho)\over q_0}\right];
\end{equation}
\begin{equation}
\label{dotrho4}
{d\tilde\rho\over d \epsilon} = {\kappa_{\rho}b^2\sigma\over \gamma_D}\,{q(\sigma,\tilde\rho)\over q_0}\,\left(1 - {\tilde\rho\over e^{-1/\tilde\chi}}\right);
\end{equation}
\begin{equation}
\label{firstlaw5}
{d\tilde\chi\over d \epsilon} = {\sigma\,\over c_{e\!f\!f} e_D}{q(\sigma,\tilde\rho)\over q_0}\,\left[1 - {\tilde\chi\over\tilde \chi_0}- \kappa_{\rho}\left(1 - {\tilde\rho\over e^{-1/\tilde\chi}}\right)\right].~~~~~~
\end{equation}

Finally, before looking at applications of this theory, note that we can simplify it by assuming that we are interested only in the plastic behavior during and after the onset of hardening.  Because $\mu$ is substantially larger than other stress scales in the problem, Eq.(\ref{dotsigma2}) is a stiff differential equation that is accurately solved by writing $q(\sigma,\tilde\rho)\cong q_0$.   Thus, Eq.(\ref{nudef}) becomes an expression for the stress:
\begin{equation}
\label{sigma-nu}
\sigma = \mu_T\,\sqrt{\tilde\rho}\,\,\nu(T,\tilde\rho,q_0).
\end{equation}
Equations (\ref{dotrho4}) and (\ref{firstlaw5}) become
\begin{equation}
\label{dotrho5}
{d\tilde\rho\over d \epsilon} = \kappa_{\rho}\,{\mu_T\,b^2\over \gamma_D}\,\sqrt{\tilde\rho}\,\,\nu(T,\tilde\rho,q_0)\,\left(1 - {\tilde\rho\over e^{-1/\tilde\chi}}\right);
\end{equation}
and
\begin{equation}
\label{firstlaw6}
{d\tilde\chi\over d \epsilon} = \kappa_{\chi}\,\sqrt{\tilde\rho}\,\,\nu(T,\tilde\rho,q_0)\left[1 - {\tilde\chi\over\tilde \chi_0}- \kappa_{\rho}\left(1 - {\tilde\rho\over e^{-1/\tilde\chi}}\right)\right],~~~~~~~~~
\end{equation}
where $\kappa_{\chi} \equiv \mu_T/ c_{e\!f\!f} e_D$ is a dimensionless number very roughly of the order of unity.

\section{Applications}

\subsection{Onset of Strain Hardening}
\label{onset}

Strain hardening is the transient approach to steady-state flow.  The physical mechanisms that determine this transient are contained in the prefactors on the right-hand sides of Eqs.(\ref{dotrho5}) and (\ref{firstlaw6}), especially in the dimensionless conversion factor $\kappa_{\rho}$.  The steady-state behavior is trivially independent of these prefactors.  As shown in LBL, using $\tilde\rho = e^{-1/\tilde\chi_0}$ -- the steady-state solution of Eq.(\ref{dotrho5}) -- in Eq.(\ref{sigma-nu}), produces temperature-dependent curves of stress {\it versus} strain rate that are in excellent agreement with experiment.  That agreement is especially striking when $\tilde\chi_0$ becomes a function of the strain rate in the strong-shock regime.  

However, it is not entirely trivial to disentangle transient from steady-state behaviors in evaluating $\kappa_{\rho}$.  To see this, look again at the onset of hardening as discussed in LBL.  Near this onset, the dislocation density is relatively small, so that the second term in the parentheses on the right-hand sides of Eqs.(\ref{dotrho4}) and (\ref{dotrho5}) is negligible. It seems plausible that the stress in this regime is simply the bare Taylor stress, i.e. $\mu$ multiplied by the strain required to move a dislocation a small fraction of an atomic spacing away from a pinning point, unmodified by the thermal effects implicit in  $\nu(T,\tilde\rho,q_0)$.  If this conjecture is correct, then Eq.(\ref{dotrho4}) becomes
\begin{equation}
\label{dotrho6}
\left({d\tilde\rho\over d \epsilon}\right)_{{\rm onset}} \cong {\kappa_{\rho}b^2\sigma_T\over \gamma_D}= {\kappa_{\rho}b^2\,\mu_T\over \gamma_D}\,\sqrt{\tilde\rho}.
\end{equation}
With no loss of generality, let $\gamma_D = \mu\,b'^2$, where $b'$ is a microscopic length, comparable to or perhaps smaller than the Burgers vector. Then
\begin{equation}
\label{hardeningrate}
{1\over \mu}\left({d\sigma\over d\epsilon}\right)_{{\rm onset}} \cong {1\over \mu}\left({d\sigma_T\over d\epsilon}\right)_{{\rm onset}}\cong {\kappa_{\rho}\,\mu_T^2\,b^2\over 2\,\mu^2\,b'^2},
\end{equation}
independent of $\tilde\rho$. According to Kocks and Mecking,\cite{KOCKS-MECKING-03} the hardening rate on the left-hand side of this equation is often found experimentally to be about $0.05$.  If $\mu_T \cong (b'/b)\,\mu$, i.e. if the length $b'$ associated with the energy per unit length of a dislocation is the same as the displacement necessary to dislodge a dislocation from a pinning site in the formula for $\sigma_T$, then $\kappa_{\rho}$ is about $0.1$. In any case, if $\kappa_{\rho}$ is a constant of order unity or less, then the predicted value of the hardening rate in Eq.(\ref{hardeningrate}) is independent of strain rate and temperature, as observed.  

Equation (\ref{hardeningrate}) is not exactly what emerges when we use the full versions of Eqs.(\ref{sigma-nu}) and (\ref{dotrho5}) to compute the onset rate.  Instead, there appears an extra factor $\nu(T,\tilde\rho,q_0)^2$ multiplying $\kappa_{\rho}$ on the right-hand side of Eq.(\ref{hardeningrate}). Thus, as in LBL, I  conclude that $\kappa_{\rho}$ is proportional to $\nu(T,\tilde\rho,q_0)^{-2}$ in Eq.(\ref{dotrho5}).  It seems to me that the physics leading to Eq.(\ref{hardeningrate}) is basically correct, as was the physics leading to the full equations in which $\kappa_{\rho}$ was an undetermined positive parameter.

\subsection{The Rate-Hardening Anomaly}
\label{upturn}

\begin{figure}[here]
\centering \epsfig{width=.5\textwidth,file=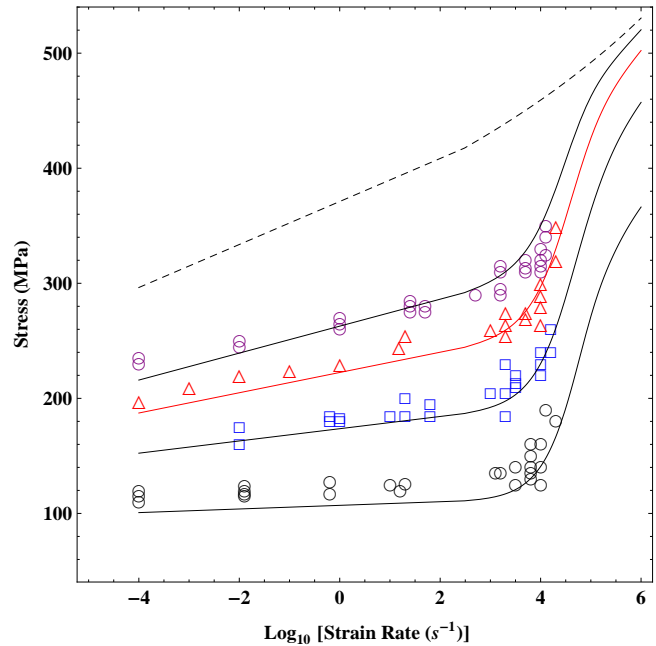} \caption{The rate-hardening anomaly as reported in  \cite{FOLLANSBEE-KOCKS-88} and \cite{PTW-03}.  The four curves, from bottom to top, show stresses as functions of strain rate for four different strains, $\epsilon = 0.05,\,0.10,\,0.15,\,{\rm and} \,0.20$. The dashed curve at the top is the theoretical steady-state prediction.} \label{SHFig1}
\end{figure}

Materials scientists have been puzzled for decades by the sudden onset of rate hardening that is seen at strain rates of the order of $10^4/{\rm sec}$, when stresses are measured at small strains. The phenomenon is illustrated in Fig.\ref{SHFig1}, where the data points (taken from \cite{FOLLANSBEE-KOCKS-88} and \cite{PTW-03}) show stresses measured in room-temperature Copper at four different strains, $\epsilon = 0.05,\,0.10,\,0.15,\,{\rm and} \,0.20$, as functions of strain rate.  

This behavior cannot be interpreted as an approach to some kind of singularity, as sometimes has been assumed in the literature.  We know with fair certainty that the  rate-hardening anomaly disappears if measured at larger strains, where the stresses must approach  apparently unremarkable steady-state values.  Strain rates of $10^4/{\rm sec}$ are about four decades smaller than those that induce rapidly growing, disorder-generated hardening, as seen in the LBL analysis of the strong-shock regime.  These two qualitatively different kinds of behavior should not be confused with each other.  

If the rate-hardening anomaly is a transient phenomenon, then its physical origin must be contained in the dimensionless prefactor $\kappa_{\rho}$.  It was assumed implicitly, in the discussion following  Eq.(\ref{hardeningrate}), that $\kappa_{\rho}$ is a conversion factor that may incorporate a wide range of physical mechanisms.  One ingredient of $\kappa_{\rho}$ must be a rate at which dislocations are created, e.g., a density of Frank-Read sources or the like, multiplied by a nucleation rate per source, presumably proportional to the stress.  To obtain a dimensionless conversion factor, multiply this rate by some time and volume associated with the driving forces.  The beauty of this approach is that we do not need yet to specify those details; we need only to know that $\kappa_{\rho}$ is a dimensionless number of the order of unity or less.

%%%%%%%%%%%%% FIGURE 1 %%%%%%%%%%%%
%%%%%%%%%%%%%%%%%%%%%%%%%%%%%%%%%%%%% 

The rate-hardening anomaly implies that $\kappa_{\rho}$ acquires a strain-rate dependence.  With the contents of the next subsection in mind, I propose that this is a grain-size effect.  Suppose that, whenever a grain of linear size $d$ is sheared by an amount $b/d$, it emits a dislocation and then relaxes back toward its original shape.  Thus, the emission frequency, multiplied by the areal density of grains, is proportional to $\dot\epsilon/d\, b$, in analogy to the source strength discussed in the preceding paragraph. This is an independent dislocation-creation mechanism that, to a first approximation, should be added linearly to the rate-independent term.   

This analysis, combined with that in Subsection \ref{onset}, suggests that we write
\begin{equation}
\label{kappanew}
\kappa_{\rho} =  {\tilde\kappa_{\rho}\over \nu(T,\tilde\rho,q_0)^2}\,\left(1 + {q_0\over q_1}\right),
\end{equation}
where $\tilde\kappa_{\rho}$ is a constant of the order of unity, and $q_1$ is a dimensionless strain rate, possibly proportional to the grain size. By writing Eq.(\ref{kappanew}) in this way, $q_1/\tau_0$ is approximately the strain rate at which the upturn occurs.  Then rewrite Eq.(\ref{dotrho5}) in the form
\begin{equation}
\label{dotrho6}
{d\tilde\rho\over d \epsilon} = \kappa_1\,{\sqrt{\tilde\rho}\over \nu(T,\tilde\rho,q_0)}\,\left(1 + {q_0\over q_1}\right)\,\left(1 - {\tilde\rho\over e^{-1/\tilde\chi}}\right),
\end{equation}
where 
\begin{equation}
\kappa_1 \equiv \tilde\kappa_{\rho}\,{b^2\,\mu_T\over \gamma_D}
\end{equation}
is a dimensionless number, again of the order of unity.  Equation (\ref{firstlaw6}) is unchanged, but the factor $\kappa_{\rho}$ inside the square brackets is now the function of $\tilde\rho$ and $q_0$ given by Eq.(\ref{kappanew}). 

Figure \ref{SHFig1} shows comparisons between experiment and theory for the rate-hardening anomaly.  The dashed black line is the theoretical steady-state curve.   The parameters used for plotting these graphs  are almost the same as those used in LBL: $T_P = 40800\,K$, $T= 298\,K$, $\mu_T = 1600\,\,{\rm GPa}$, $\chi_0 = 0.25$, $\kappa_{\chi} = 16$, and $\kappa_1 = 3.1$ (the ``universal'' Kocks-Mecking value based on  observed onset rates). This upturn analysis seems to be insensitive to the storage term in Eq.(\ref{firstlaw6}); therefore I have omitted it here by setting $\kappa_{\rho} = 0$ in that equation. (That term will be nonzero in the next subsection.) The rate defined in Eq.(\ref{kappanew}) is $q_1/\tau_0 = 4 \times 10^4\,{\rm sec}^{-1}$, with $\tau_0 = 10^{-12}$ seconds.  The initial values of $\tilde\rho$ and $\tilde\chi$ used for integrating the differential equations are $\tilde\rho_i = 10^{-7}$ and $\tilde\chi_i = 0.18$, again the same as in LBL. 

Note that $q_1$ is the single new parameter that is needed to explain the rate-hardening anomaly for all four values of the strain shown here.  The data for $\epsilon = 0.15$ is the least noisy of these data sets, and shows the cleanest upturn.  It is this curve that appears most often in the literature. The agreement for the outlying cases, $\epsilon = 0.05$ and $0.20$ is somewhat less good, especially at the  upturns where the data is most noisy.  But the overall agreement between theory and experiment seems excellent.

\subsection{Effects of Grain Size}
\label{grainsize}

As a second example of the statistical thermodynamics of dislocations, consider the effects of grain size in room-temperature Copper as observed by Meyers et al.\cite{MEYERSetal-95}  That paper is largely devoted to measurements of grain-size effects in phenomena such as dynamic recrystallization and strain localization.  But it contains, in its Figs. 5 and 6, some data that is directly related to the present investigation.  These figures show that the Kocks-Mecking rule of constant onset rate is violated at small grain sizes.  The initial slopes of the stress-strain curves increase markedly with decreasing grain size. Moreover, at the smaller grain sizes, these curves exhibit narrow transitions  between rapid onsets at small strains and slower hardening at larger strains.  The present theory can account for most of this behavior.  

%%%%%%%%%%%%% FIGURE 2 %%%%%%%%%%%%
\begin{figure}[here]
\centering \epsfig{width=.5\textwidth,file=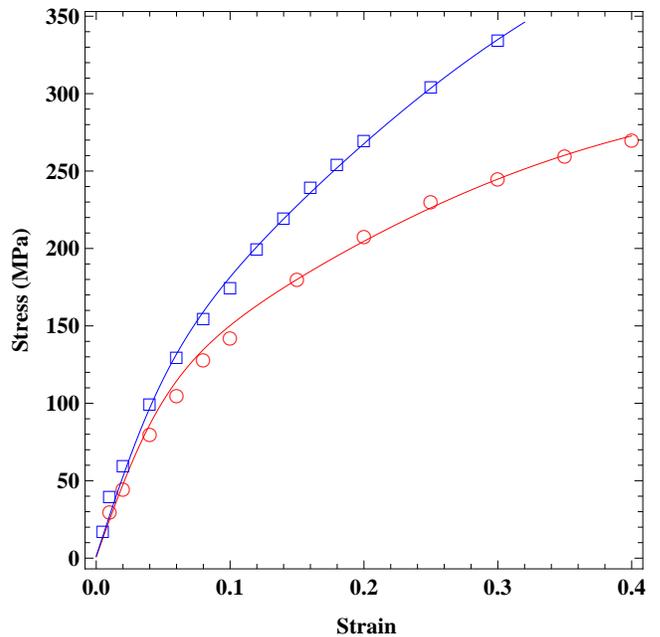} \caption{Stress-strain curves for the largest grain size, $d = 315\, \mu{\rm m}$ . The strain rates are $10^{-3}$ and $3\,\times 10^3\,{\rm sec}^{-1}$ for the bottom and top curves respectively. The data points are taken from \cite{MEYERSetal-95}.} \label{SHFig2}  
\end{figure}
%%%%%%%%%%%%%%%%%%%%%%%%%%%%%%%%%%%%% 

%%%%%%%%%%%%% FIGURE 3 %%%%%%%%%%%%
\begin{figure}[here]
\centering \epsfig{width=.5\textwidth,file=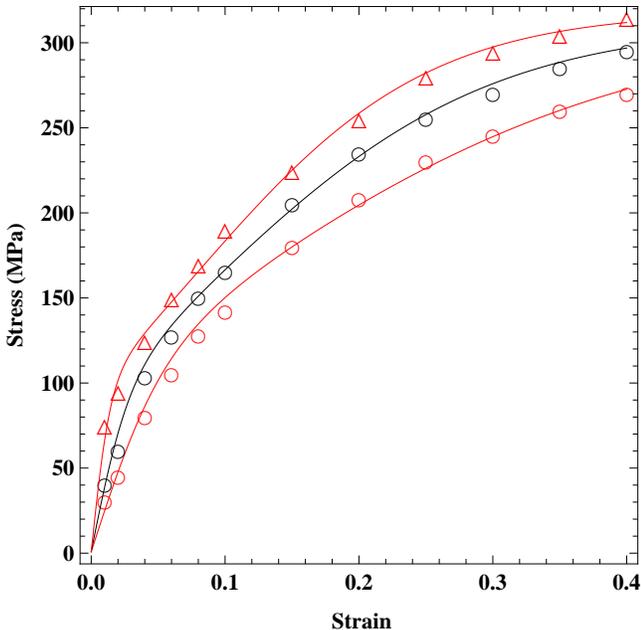} \caption{Stress-strain curves at the strain rate $10^{-3}\,{\rm sec}^{-1}$, for different grain sizes: $d = 9.5,\,117$, and $315$ $\mu{\rm m}$, from top to bottom. The data points are taken from \cite{MEYERSetal-95}.} \label{SHFig3}
\end{figure}
%%%%%%%%%%%%%%%%%%%%%%%%%%%%%%%%%%%%% 

Meyers et al. show stress-strain curves for materials with four different grain diameters: $d = 9.5,\,25,\,117$, and $315$ $\mu{\rm m}$. For each $d$, they show curves for strain rates $10^{-3}$ and $3\,\times 10^3\,{\rm sec}^{-1}$. To establish a base-line for the present analysis, start with the largest grain size, $d = 315$ $\mu{\rm m}$, where there should be little, if any, anomalous strain-rate dependence.  The theoretical results are shown in Fig.\ref{SHFig2} along with experimental points taken from \cite{MEYERSetal-95}.  The theoretical parameters are essentially the same as those used in Sec. \ref{upturn}.  The only differences are that $q_1/\tau_0 = 3 \times 10^4\,{\rm sec}^{-1}$ (slightly smaller than before); and that $\kappa_{\chi} = 12$ and $8.5$ for the smaller and larger strain rates respectively. Note that the onset rate is quantitatively as predicted by Kocks and Mecking, i.e. $\kappa_1 = 3.1$, so that both curves have the same initial slope although their strain rates differ by more than six orders of magnitude.

\begin{figure}[here]
\centering \epsfig{width=.5\textwidth,file=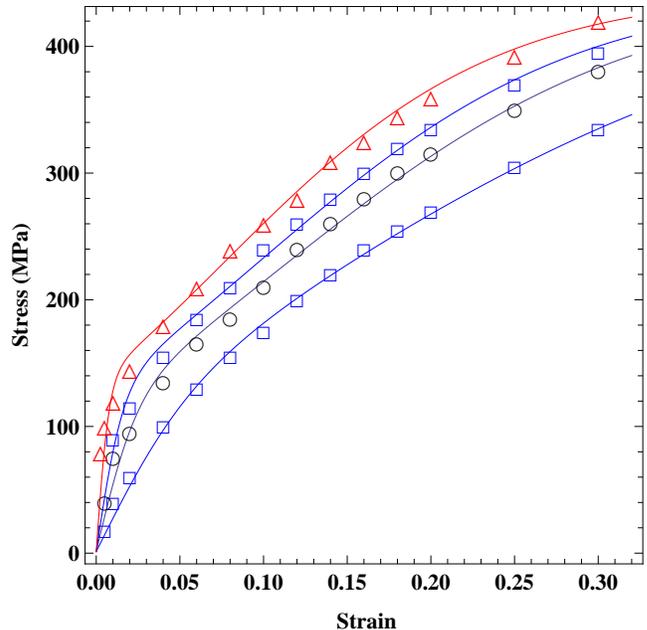} \caption{Stress-strain curves at the strain rate $3 \times 10^{3}\,{\rm sec}^{-1}$, for different grain sizes: $d = 9.5,\,25,\,117$, and $315$ $\mu{\rm m}$, from top to bottom. The data points are taken from \cite{MEYERSetal-95}. } \label{SHFig4}
\end{figure}

This situation changes dramatically when we look at the smaller grain sizes as shown in Fig.\ref{SHFig3}.  Here, I have plotted stress-strain curves for three grain sizes, $d = 9.5,\,117$, and $315$ $\mu{\rm m}$, all at the lower strain rate, $q_0/\tau_0 = 10^{-3}$, where the rate-dependent term $q_0/q_1$ in Eq.(\ref{kappanew}) should be negligible. (I have omitted the intermediate curve for $d = 25\,\mu{\rm m}$ for the sake of clarity.)  The lower curve in this figure, for $d = 315\,\mu{\rm m}$, is the same as the lower curve in Fig.\ref{SHFig2}.  However, the upper two curves have substantially greater onset slopes, which means that the energy-conversion factor $\kappa_{\rho}$ or, equivalently, $\kappa_1$, must be a function of the grain size.  The density of ordinary dislocation sources -- not only the  strain-rate driven ones discussed earlier -- apparently increases with the density of grain boundaries.  Accordingly, I have used the small strain-rate data in Fig.\ref{SHFig3} to evaluate $\kappa_1$ as a function of $d$.  I then have used those values to compute the stress-strain curves at the higher strain rate, $q_0/\tau_0 = 3 \times 10^{3}\,{\rm sec}^{-1}$ shown in Fig.\ref{SHFig4}. 
 
The theoretical curves shown in Figs. \ref{SHFig3} and \ref{SHFig4} are uncertain in many respects, none of which seem to be fatal to the main concepts being tested here.  In the first place, the data shown in  \cite{MEYERSetal-95} is noisy, and my own ability to extract accurate points from it is limited.  Secondly, there are too many things happening in this theory and, thus, too many parameters to evaluate.  My strategy has been to start by fixing a few parameters based on prior experience with Cu data, e.g. in LBL and in the upturn analysis reported here, despite the fact that the materials used by Meyers et al are not exactly the same as those used elsewhere.  Thus, I again use  $T_P = 40800\,K$, $T= 298\,K$, $\mu_T = 1600\,\,{\rm GPa}$, and $\tilde\chi_0 = 0.25$. The initial values of $\tilde\rho$ and $\tilde\chi$ are again $\tilde\rho_i = 10^{-7}$ and $\tilde\chi_i \cong 0.18$, except that I have had to make small adjustments of $\tilde\chi_i$ as noted below. 

The storage term, indicated by the factor $\kappa_{\rho}$ inside the square brackets in Eq.(\ref{firstlaw6}), may be playing a role here.  This is a thermodynamically predicted softening effect.  The rate at which energy is stored in dislocations reduces the rate at which the entropy of disorder is increasing, thus reducing the rate at which the density of dislocations is increasing and, in turn, reducing the hardening rate.  That mechanism seems to be playing a role here in reducing the slope of the stress-strain curves after the initial rise at small grain sizes. I have tentatively accounted for it by letting $\tilde\kappa_{\rho} = 0.2$ in the formula for $\kappa_{\rho}$ in  Eq.(\ref{kappanew}) for all grain sizes and strain rates.  

The values of the onset parameter determined at small strain rates, for $d = 9.5,\,25,\,117,\,{\rm and}\,\,315\,\mu{\rm m}$, are $\kappa_1 = 9.5,\,7.5,\,5.0,\,{\rm and}\,3.1$ respectively.  The corresponding values of $q_1/\tau_0$, needed for the high strain-rate analysis, are $2\times 10^3,\,6 \times 10^3,\,7 \times 10^3,\,{\rm and}\,\,3 \times 10^4\,{\rm sec}^{-1}$.  Thus, as expected, both the ordinary conversion factor and the strain-rate induced one increase as the grain size decreases.  

%%%%%%%%%%%%% FIGURE 4 %%%%%%%%%%%%
%%%%%%%%%%%%%%%%%%%%%%%%%%%%%%%%%%%%% 

To fit the experimental data to the accuracy shown in the figures, I have had to make relatively small adjustments of other parameters.  For the same increasing sequence of grain sizes listed in the preceding paragraph, I find $\kappa_{\chi} = 18,\,17,\,17,\,12$ for the small strain rate, and $13,\,11.5,\,11.5,\,8.5$ for the large one.  The corresponding values of $\tilde\chi_i$ are $0.165,\,0.165,\,0.17,\,0.18$, and $0.16,\,0.16,\,0.16,\,0.18$.  For $d = 9.5\,\mu{\rm m}$, the small strain-rate curve shown in Fig.\ref{SHFig3} uses  $\tilde\kappa_{\rho} = 0.4$ and $\tilde\chi_0 = 0.253$.  So far as I can tell, these final adjustments serve only cosmetic purposes.  Neither the experimental data nor the theoretical analysis are accurate enough for us to draw more definite conclusions from them.

\section{Closing Remarks}

The analysis presented here contains several unconventional predictions that might usefully be checked experimentally.  Specifically, it predicts that the anomalous rate-hardening curves level off at higher strain rates, and that the anomaly disappears at larger strains. It also predicts that the upturn should occur at smaller strain rates for smaller grain sizes, and even speculates a simple power law dependence for this effect. In other words, it points toward extensions of earlier experiments.  

I close by repeating the plea from the conclusion of LBL.  The theory presented there, and here, should be useful as a basis for studying dynamic, spatially heterogeneous instabilities such as strain localization.  If dislocation theory is to serve as a realistic tool for predicting the strength of materials, it must move in that direction.  The thermodyamic STZ theory is developing successfully in that way as seen, for example, in its prediction of the fracture toughness of bulk metallic glasses.\cite{RYCROFT-EB-12}  I see no reason why the present dislocation theory cannot be used similarly.   

\begin{acknowledgments}

This paper was motivated by discussions at the program on ``Avalanches, Intermittency, and Nonlinear Response in Far-from-Equilibrium Solids'' at the Kavli Institute for Theoretical Physics, University of California, Santa Barbara, September - December 2014.  I am especially grateful to Sid Yip for reawakening my interest in the dislocation problems.  This research  was supported in part by the U.S. Department of Energy, Office of Basic Energy Sciences, Materials Science and Engineering Division, DE-AC05-00OR-22725, through a subcontract from Oak Ridge National Laboratory.

\end{acknowledgments}

\end{document}